\documentstyle{aipproc}

\begin{document}
\title{The Transverse Quark Distribution and Proton
Electromagnetic Form Factors in Skew Distribution Formalism}

\author{John P. Ralston$^a$, Pankaj Jain$^b$ and Roman V. Buniy$^a$}
\address{
$^a$ Department of Physics and Astronomy, University of Kansas, Lawerence,
KS 66045, USA\\
$^b$ Physics Department, I.I.T. Kanpur, India 208016
}
\maketitle

\begin{abstract} Skew density matrices can be diagonalized to yield probability
interpretation.  The power-counting prediction of perturbative QCD is
found consistent with recent CEBAF data on $F_2(Q^2)/F_1(Q^2)$.

\end{abstract}

Perturbative $QCD$ can be applied to hard exclusive processes in many
ways.  Use of ``skew" or ``off-diagonal" parton distributions
\cite{skew} has generated attention.  Skew distributions are matrix
elements sharing features with density matrices.  The diagonal
elements of a positive density matrix are interpreted as
probabilities.  Density matrices depend on the basis: probability in
one frame will appear to be an interference of amplitudes in another
frame.  This occurs with quark
spin amplitudes\cite {Ralston79}, where transverse polarization 
(``transversity")
distributions appear as interference in the helicity basis, and
helicity distributions appear as interference in the transverse
polarization basis.

Here we find a probabilistic interpretation for skew distributions
involved in the proton electromagnetic form factors.  Then a simple,
kinematic angular momentum argument resolve a recent puzzle from data
for the electromagnetic form factors $F_1(Q^2), F_2(Q^2)$.  For about
30 years it was thought that $F_2(Q^2)/F_1(Q^2)\sim 1/Q^2$ was a
prediction of hard scattering, and in particular, of $pQCD$.  We made the
same prediction on the basis of skew distributions in
1993\cite{jainral93}.  Recent $CEBAF$ data\cite{jonesPRL} shows that
$F_2/F_1 \sim 1/\sqrt{|Q^2|}.  $ Our prediction missed some
instructive points, and the $CEBAF$ result is expected, if proper
kinematics is simply taken into account.

{\it Diagonalization:}  Consider $2 \rightarrow 2$ quark-proton 
scattering with momentum transfer
$\Delta^\mu, \Delta^2<0$.  The proton is described by states $ |p-\Delta/2,
s\! > $ coming ``in" and $|p+\Delta/2, s'\! >$ going ``out". The 
scattering matrix element is
\begin{eqnarray}
&& \Phi_{\alpha, \beta}^{s,s'}(p; k, k',
\Delta)\delta(k+p-k'-p-\Delta) \nonumber \\ &&= \int\!  d^4 z d^4 z'\,
\exp(-ikz +ik'z') <\!  p-\Delta/2, s | \psi_{\beta}(z) \bar
\psi_{\alpha}(z') |p+\Delta/2, s'\!  >.  \nonumber \end{eqnarray} Here
$\alpha, \beta$ are Dirac indices of the quark fields $\psi$, and the
in- and out-quarks have momenta $k, k'$ with $k=x (p-\Delta/2) +\vec
k_T, k'=x'(p+\Delta/2)+\vec k_T'$.  A time-ordering symbol is dropped,
as only one time ordering contributes: see Diehl and Gousset\cite
{skew} Now $\Phi (\Delta)$ can be decomposed into terms symmetric and
antisymmetric in $\Delta$; the symmetric part is Hermitian.  This
leads to a density matrix we can diagonalize.

To diagonalize we make a series of coordinate transformations.  Make a
Lorentz transformation to the frame: $p = ( p^+, 0, m^2/2 p^+);\ \
\Delta = ( 0, \Delta_T, -t/2 p^+);\ \ \Delta_T^2=-t.  $ Now
$\Delta^{\mu}$ becomes entirely transverse as $p^+ \rightarrow
\infty$.  The partons have 4-vectors $ k^{\mu}= xp^{\mu} + k_T^{\mu},
x( p+ \Delta)^{\nu} + k_T'^{\nu}$ , yielding $$x=x';\ \ k_T' = k_T +
\Delta_T (1-x).$$ The matrix element is now diagonal in $x$.  This
might seem impossible, because the $x,x'$ dependence of skew
distributions is thought to be invariant: but $x$ is not
Lorentz-invariant sideways.  The convolution in $\vec k_T$ is
diagonalized by conjugate transverse spatial coordinates $\vec b$.
Then $ \int\!  d^2 b \ \Phi(s,s'; \vec b/(1-x), x) e^{i \vec b \cdot
\vec \Delta}$ can be inverted by Fourier transform to find the
integrand, diagonal in everything but spin, which also can be made
diagonal by familiar helicity or transverse bases.

The choice of frames and diagonalization was used in an independent
early introduction of off-diagonal distributions\cite{jainral93}.  Due
to a kinematics goof we missed the factor $1/(1-x) $.  A review by
Brodsky and Lepage \cite{Brodsky89} was useful.

{\it Interpretation:}  The electromagnetic form factors (with 
$-Q^2=\Delta_T^2$) are found
by following the
Feynman rules, which includes multiplication by the quark charge $e_q$,
tracing $\Phi$ with $\gamma^{\mu}  $, and
doing the integral $\int\! dx\ d^2 b$.  At large $-Q^2$ this is dominated by
$b^2
\sim 1/|Q^2| $ by Fourier analysis.  The ``short distance" implied by
large $Q$ is far more general than the more problematic
``quark-counting" argument.  We simply have {\it one} quark
located by the hard momentum $Q$, while the asymptotic short distance
theory\cite{Brodsky89} assumes that {\it all possible Fock states} 
are separated by
asymptotically short distance.  We remain within the framework of
$pQCD$, of course, while choosing a more general factorization method.

Since the form factor $F_1(Q^2)$ is known from data, we invert the Fourier
transform to solve for a positive-definite diagonal
element of the density matrix, namely a probability:
\begin{equation}
P(\lambda,\lambda;
\vec b)=\frac{ 1 }{ e_q^{2} } \int\! d^2 Q_T\ e^{-i \vec b \cdot \vec Q_T}
F_1(Q^2).
\end{equation}
Positivity is assured because $F_1(Q^2)$ is monotonically
decreasing.  The quantity $P(\lambda,\lambda;
\vec b)$ is on a similar footing to the usual parton distributions.
Indeed, the usual parton distributions are functions of $x$
integrated over $k_T$, while $P(b)$ depends on $b=|
\vec b|$ and has been integrated over $x$.

A fit to the Fourier transform of $F_1(Q^2)$ has been performed.  A profile
of the transverse ($1/(1-x) $-weighted) probability to find a quark (mostly
of the up-type) was shown at the meeting.  We expect that this 
invariant quantity will be useful in many
studies involving the transverse coordinate.

{\it Orbital Angular Momentum: } Quark orbital angular momentum is a
fascinating subject of great interest in the proton spin puzzle.  In
the case at hand, we are lucky to have two quarks evaluated at the
same $x, \vec b$ points, so that difficulties of gauge invariance are
minimal.  Indeed, $P(b)$ is gauge invariant by definition in terms of
observable quantities.

We expand the operators in terms of solutions to $\nabla^2
\psi=0$. By usual methods \cite{Ralston79} the operators are evaluated
inside the proton state, letting the correlation $\Phi$ be expressed by
$c$-number ``wave functions".  Partons are below threshold in a form factor,
so we need an expansion for orbital angular momentum for spacelike $k$:
\begin{equation}
\psi (z^-, \vec b ,z^+ =0)  = \sum_n \int\! dx\ \psi_n(x) I_n( b) e^{- i n
\phi} e^{i xp^+ z^-} .
\end{equation}
Here $e^{- i n \phi}$ are light-cone $SO(2)$ orbital angular momentum
basis functions; $I_n( b)$ are modified Bessel functions more usually
seen as $J_n (b)$ for a timelike basis.  \footnote { If one is
concerned about $b \rightarrow \infty, $ then Green functions can be
expanded in series of $I_n(b_<) K_n(b_>)$ where $b_< (b_>)$ is the
smaller (larger) of two $b$ arguments.  We only need short distance.
} We are interested in the $F_2$ form factor, associated with
$i\sigma_{\mu \nu} Q^{\nu}/2m$, which represents proton helicity-flip
at large $Q^2$.  Since we cannot change the helicity of a quark with a
hard scattering (the near-perfect chiral symmetry of $pQCD$), the
proton can only flip its spin to make $F_2$ by transferring a unit of
{\it orbital } angular momentum in the quark \cite{jainral93}.
(Hoodbhoy and Ji \cite{Hoodbhoy} subsequently verified the same
result.)

Note we are not attempting here to derive the functional dependence of
$P(\lambda, \lambda;\vec b)$: in our approach this comes from data.
Our approach is to relate each power of $b$ or angular momentum to
further suppression by powers of $1/Q$.  The argument for short
distance is kinematically compelling here (if controversial in the
quark-counting method), so for large $Q$ the Fourier transform is
dominated by $I_0(b) \sim b^0$ if this channel is allowed.  But
$I_0(b)$ represents the s-wave component, with zero angular momentum,
and so this channel is open only to the helicity non-flip, namely
\begin{equation}F_{1 }\sim\delta_{ \lambda \lambda'}
\int\! dx\ d^2b\  \bar\psi_{0}(x) \Gamma \psi_0(x)
\left[I_{0}\left( b/(1-x) \right)\right]^2 \exp(i \vec Q_T\cdot \vec 
b).  \end{equation}
Here $\Gamma $ represents the
necessary Dirac matrices.

Conversely, the only possibility for the helicity-flip $F_2$ is to use
powers of $b$ to conserve the angular momentum, and suffer the
corresponding power-suppression in $Q$.  On the basis of this power
counting, it was reasoned \cite {jainral93} that by angular
momentum selection rules, the integrals over $b$ would vanish unless
two representations of the same angular momentum matched up, giving
the previous prediction $F_2(Q^2)/F_1(Q^2) \sim 1/Q^2$ for $Q^2>
GeV^2$.  Now consulting the formulas this is simply not true.  There
is a factor of $e^{ i \vec Q_T\cdot \vec b_T} $ carrying angular
momentum:
\begin{equation}
F_{2 }\sim \delta_{ \lambda,- \lambda'}\int\!  dx\ d^2b\
\bar\psi_{0}(x) \Gamma \psi_1(x) I_{0}\left( b/(1-x)
\right)I_{1}\left( b/(1-x)\right ) e^{ i |Q_T| b\cos(\phi-\phi_Q)
}e^{i\phi} + cc.\end{equation} Physically, the probe $\vec Q$ breaks
the rotational symmetry of the problem.

We reiterate that this analysis is entirely within the context of
$pQCD$.  In $pQCD$ one takes some matrix elements from the data, and
makes predictions for others.  What can we predict here?  From the
power-counting cited, we have
\begin{equation}
{F_2(Q^2)\over F_1(Q^2)} = \frac{ <\! \bar \psi_1(x)\psi_0(x)\!>}{\sqrt{|Q^2|}
<\! b \bar
\psi_0(x)\psi_0(x)\!>},
\end{equation}
where the braces represent the integrals.  The fact that $(Q/GeV)
F_2(Q^2)/F_1(Q^2) $ is not far from unity in the CEBAF data indicates
that the proton wave functions for quark angular momenta 1 and 0 are
not too different in magnitude.  Constituent quark, or
non-relativistic models are ruled out, but those models were never
capable of capturing the Fock space description of the skew
distribution.  It will be interesting to continue these studies in the
context of the larger proton spin puzzle.

\bigskip \noindent {\bf Acknowledgments:} Work supported by DOE grant
number DE-FGO2-98ER41079.

\end{document}